\documentclass[a4paper]{article}

\usepackage{amsmath,amssymb,graphicx,wrapfig,RR,hyperref}
\usepackage{pslatex,color,bbm}
\usepackage{algorithm}
\usepackage[noend]{algorithmic}
\usepackage{xspace}
\usepackage{lineno}

 \newtheorem{theorem}{Theorem}

 \newtheorem{definition}{Definition}
\newtheorem{corollary}{Corollary}
 \newtheorem{lemma}{Lemma}
\newtheorem{newproperty}{Property}

\newenvironment{remark}[1][]{\par \noindent {\bf Remark #1}\ }{\par\vspace{11pt}}

\newenvironment{proof}[1][]{\par \noindent {\bf Proof #1}\ }{\hfill$\Box$\par \vspace{11pt}}

\newcommand{\pn}{\ensuremath{\rm pn}}
\newcommand{\uu}{\ensuremath{\rm pn}}
\newcommand{\vv}{\ensuremath{\rm pn^+}}
\newcommand{\uuu}{\ensuremath{\rm pn}}

\newcommand{\pw}{\ensuremath{{\rm pw}}}

\newcommand{\vs}{\ensuremath{{\rm vs}}}
\newcommand{\ns}{\ensuremath{{\rm ns}}}
\newcommand{\sn}{\ensuremath{{\rm ns}}}
\newcommand{\es}{\ensuremath{{\rm es}}}
\newcommand{\N}{\ensuremath{{\mathbbm N}}}
\newcommand{\set}[1]{\ensuremath{\left\{#1\right\}}\xspace}
\def\AT{\ensuremath{aT}}
\def\algoHD{\texttt{algoHD}\xspace}
\def\incHD{\texttt{IncHD}\xspace}

\RRdate{June 2008}

\RRNo{6560}

\RRtitle{Un algorithme distribu\'e pour le calcul et la mise \`a jour du process number d'une for\^et}

\RRetitle{A distributed algorithm for computing and updating the process
  number of a forest}

\titlehead{A distributed algorithm for the process number of a forest}

\RRauthor{David Coudert\thanks[1]{This work was partially funded by the European projects {\sc ist fet Aeolus} and COST 293 {\sc Graal}, and done within the {\sc crc Corso} with France Telecom R\&D.} \and  Florian Huc\thanksref{1} \and Dorian Mazauric\thanksref{1}} 

\authorhead{Coudert, Huc, Mazauric}

\RRnote{MASCOTTE, INRIA, I3S(CNRS/UNSA), Sophia Antipolis, France.\\
\texttt{\{firstname.lastname@sophia.inria.fr\}}}

\graphicspath{}

\RRresume{Dans cet article, nous pr\'esentons un algorithme distribu\'e permettant de calculer divers param\`etres d'un arbre tel le process number, la pathwidth et l'edge search number. Cet algorithme n\'ecessite $n$ \'etapes, a un temps d'ex\'ecution de $O(n\log{n})$ et g\'en\`ere $n$ messages de taille $\log_3{n}+3$. Nous montrons ensuite comment il peut servir a mettre \`a jour le process number (ou la pathwidth ou l'edge search number) de chaque composante d'un for\^et apr\`es l'ajout ou la suppression d'une ar\^ete. En fin on montre que cela peut \^etre fait m\^eme si la taille de la for\^et est inconnue.}

\RRabstract{
  In this paper, we present a distributed algorithm to compute various
  parameters of a tree such as the process number, the edge search
  number or the node search number and so the pathwidth. This
  algorithm requires $n$ steps, an overall computation time of
  $O(n\log{n})$, and $n$ messages of size $\log_3{n}+3$. We then
  propose a distributed algorithm to update the process number (or the
  node search number, or the edge search number) of each component of
  a forest after adding or deleting an edge. This second algorithm
  requires $O(D)$ steps, an overall computation time of $O(D\log{n})$,
  and $O(D)$ messages of size $\log_3{n}+3$, where $D$ is the diameter
  of the modified connected component.\linebreak Finally, we show how to extend
  our algorithms to trees and forests of unknown size using messages
  of less than $2\alpha+4+\varepsilon$ bits, where $\alpha$ is the
  parameter to be
  determined and $\varepsilon=1$ for updates algorithms.\\
}

\RRkeyword{pathwidth, process number, search number, distributed algorithm.}
\RRmotcle{pathwidth, process number, search number, algorithme distribu\'e}

\URSophia
\RRprojet{MASCOTTE}
\RRtheme{\THCom}

\begin{document}

\makeRR 

\section{Introduction}
Treewidth and pathwidth have been introduced by Robertson and
Seymour~\cite{RoSe83} as part of the graph minor project.  By
definition, the treewidth of a tree is one, but its pathwidth might be
up to $\log{n}$. A linear time centralized algorithms to compute the
pathwidth of a tree has been proposed in~\cite{EST94,Sch90,Sko03}, but
so far no dynamic algorithm exists.

The algorithmic counter part of the notion of pathwidth is the node
searching problem~\cite{KP86}. It consists in finding an invisible and
fast fugitive in a graph using the smallest set of agents. The minimun
number of agents needed gives the pathwidth. Other graph invariants
closely related to the notion of pathwidth have been proposed
such as the process number~\cite{CPPS05,CoSe07} and the edge search
number~\cite{MHG+88}. For this two invariants it is not known if they
are strictly equivalent to the pathwidth or not.

In this paper, we propose a dynamic algorithm to compute those
different parameters on trees and to update them in a forest after the
addition or deletion of an edge. We also show that no distributed
algorithm can always transmit a number of bits linear in $n$ and give
a characterisation of the trees whose process number and edge search
number equals their pathwidth.  To present our results, we concentrate
on the process number.

As mentioned before the process number of a (di)graph has been
introduced to model a routing reconfiguration problem in WDM or WiFi
networks in~\cite{CPPS05,CoSe07}. The graph represents a set of tasks
that have to be realized.  A \emph{process strategy} is a serie of
actions in order to realize all the tasks represented by the graph. It
finishes when all the nodes of the graph are \emph{processed}. In
order to process the graph, the three actions we can do are:
\begin{itemize}
\item[(1)] put an agent on a node.
\item[(2)] remove an agent from a node if all its neighbors are either
  processed or occupied by an agent. The node is now processed.
\item[(3)] process a node if all its neighbors are occupied by an
  agent (the node is surrounded).
\end{itemize}
A \emph{$p$-process strategy} is a strategy which process the graph
using $p$ agents. The \emph{process number} of a graph $G$, $\uu(G)$,
is the smallest $p$ such that a $p$-process strategy exists. For
example, a star has process number 1 (we place an agent on its
center), a path of length at least 4 has process number 2, a cycle of
size 5 or more has process number 3, and a $n\times n$ grid has
process number $n+1$. Moreover, it has been proved
in~\cite{CPPS05,CoSe07} that $\pw(G)\leq \pn(G)\leq \pw(G)+1$, where
$\pw(G)$ is the \emph{pathwidth} of $G$~\cite{RoSe83}.

The node search number~\cite{KP86}, $\sn(G)$, can be defined similarly
except that we only use rules (1) and (2).  It was proved by
Ellis \emph{et al.}~\cite{EST94} that $\sn(G)=\pw(G)+1$, and by
Kinnersley~\cite{Kin92} that $\pw(G)=\vs(G)$, where $\vs(G)$ is the
\emph{vertex separation} of $G$. Those results show that the vertex
separation, the node search number and the pathwidth are
equivalent. Please refer to recent surveys~\cite{FT,DPS02} for more
information.



The following Theorem gives a construction which enforces each
parameter to grow by~$1$, which implies that for any tree $\ns(T)$,
$\es(T)$, $\pw(T)$, $\vs(T)$, and $\pn(T)$ are less than $\log_3(n)$.

\begin{theorem}[\cite{CPPS05} and \cite{Par76}]\label{p3}
  Let $G_1, G_2$ and $G_3$ be three connected graphs such that
  $\vs(G_i)=vs$, $\sn(G_i)=ns$ and $\pn(G_i)=p$, $1\leq i\leq 3$. We
  construct the graph $G$ by putting one copy of each of the $G_i$,
  and we add one node $v$ that has exactly one neighbour in each of
  the $G_i$, $1\leq i\leq 3$.  Then $\vs(G)=vs+1$, $\sn(G)=ns+1$ and
  $\pn(G)=p+1$.
\end{theorem}


The algorithm we propose is based on the decomposition of a tree into
subtrees forming a \emph{hierarchical decomposition}. It is fully
distributed, can be executed in an asynchronous environment and the
construction of the hierarchical decomposition requires only a small
amount of information.

It uses ideas similar to the ones used by Ellis \emph{et
  al.}~\cite{EST94} to design an algorithm which computes the node
search number in linear time. However their algorithm is centralized
and the distributed version uses $O(n\log{n})$ operations and transmit
a total of $O(n\log{n}\log(\log{n}))$ bits.  We improve the
distributed version as our algorithm also requires $O(n\log{n})$
operations but transmit at most $n(\log_3{n}+3)$ bits. We also prove
that it is optimal in the sense that for any $k \in \N$, no dynamic
algorithm, such that the vertex at which the edge addition/deletion is
done, can only simultaneously sends one message to its neighbours, can
always transmit less than $\frac{k-1}{k}n(\log_3(n))$ bits.
Furthermore, with a small increase in the amount of transmitted
information, we extend our algorithm to a fully dynamic algorithm
allowing to add and remove edges even if the total size of the tree is
unknown.

Finally we explain how to adapt our algorithm to compute the node
search number and the edge search number of a tree. It should also
certainly be adapted to compute the mixed search number and other
similar parameters.

This paper start with the presentation of the hierarchical
decomposition of a tree in Section~\ref{sec:hd}. Then in
Section~\ref{sec:algodirect} we present an algorithm to compute the
process number of a tree and analyze its complexity. In
Section~\ref{sec:incremental} we show how to update efficiently the
process number of each component of a forest after the addition or the
deletion of any tree edge, thus resulting in a dynamic algorithm.
Section~\ref{sec:discussion} concludes this paper with several
improvements including extensions of our algorithm to trees of unknown
size and to compute other parameters.

All along this paper, we assume that each node $u$ knows the set of
its neighbours which we note $\Gamma(u)$. However, the size of the
tree is not needed as explained in Section~\ref{sec:discussion}.

\section{Tools for the algorithm}
\label{sec:hd}

The algorithm is initialized at the leaves. Each leaf sends a message
to its only neighbor which becomes its {\it father}. Then, a node $v$
which has received messages from all its neighbors but one process
them and sends a message to its last neighbor, its {\it father}. We
say that this node has been \emph{visited}. Finally, the last node,
$w$, receives a message from all its neighbours and computes the
process number of $T$: $\pn(T)$. $w$ is called the {\it root} of $T$.


Notice that our algorithm is fully distributed, that it can be
executed in an asynchronous environment (we assume that each
node knows its neighbors) and that there are as many steps as nodes in
the tree.

\medskip

\begin{wrapfigure}[9]{r}{3.7cm}
   \vspace{-0.8cm}
  \begin{center}
  \scalebox{0.4}{\input{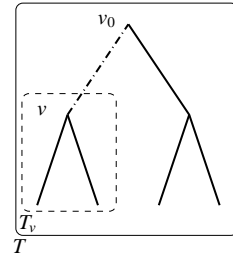}}
  \vspace{-0.3cm}
  \caption{The subtree $T_v$}
  \label{figx:Tv}
  \end{center}
\end{wrapfigure}

At each step, the goal of the message sent by a node $v$ to its father
$v_0$ is to describe, in a synthetic way, the structure of the
\emph{subtree $T_v$ rooted at $v$}, that is the connected component of
$T$ minus the edge $vv_0$, ($T - vv_0$), containing $v$ (see
Figure~\ref{figx:Tv}).

In fact a message describes a decomposition of $T_v$ into a set of
smaller disjoint trees. The trees of this decomposition are indexed by
their roots; we note $R_v$ the set of roots of the trees of this
decomposition. Through the algorithm, given a node $w$, an unique tree
with root $w$ will be computed, i.e. if in two different
decompositions there is a tree rooted at $w$, it will be the same.  We
call a tree of a decomposition with root $w$ an {\it associated-tree}
and note it $\AT^w$.

An associated-tree, and more generally any tree, can be of two types:
\emph{stable} or \emph{unstable}. Intuitively, the process number of a
stable tree will not be affected if we add a component of same process
number whereas the process number of an unstable tree will increase in
this case.

\begin{definition}\label{def:stable}
  Let $T$ be a tree with root $r$. $T$ is said \emph{stable} if there
  is an optimal process strategy such that the last (or equivalently
  first) node to have an agent is $r$ or if there is a $(\leq
  2)$-process strategy finishing with $r$. Otherwise $T$ is
  \emph{unstable}. The node $r$ is said stable or unstable accordingly
  to $T$.
\end{definition}

\begin{remark}
  We consider a tree of process number one as stable (even if an
  optimal process strategy finishing at its root needs two agents) for
  technical reason.
\end{remark}

\begin{figure}[t]
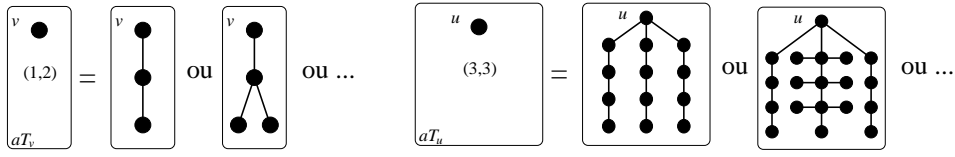

  \begin{center}
    \hfill
    \scalebox{0.33}{\input{12.pstex_t}}\hfill
    \scalebox{0.33}{\input{33.pstex_t}}\hfill\ 
    \caption{Example of trees whose associated vector are
      $vect(v)=(1,2)$ and $vect(u)=(3,3)$.}
    \label{fig:arbre}
  \end{center}
\end{figure}

From Definition~\ref{def:stable}, we give two values to describe if an
associated-tree $\AT^w$ rooted at $w$ is stable or unstable and to
give its process number: $\uu$ its process number, and $\vv$ the
minimun number of agents used in a process strategy such that the last
(or first) node to have an agent is $w$.  They together formed the
vector associated to $\AT^w$: $vect(w)=(\uu,\vv)$. By extension we
associate $vect(w)$ to $w$.  Remark that they are unique for a given
associated-tree but several associated-trees can have the same values,
also they depend on the root of the associated-tree (see
Figure~\ref{fig:arbre}). Remark also that to store this vector it is
sufficient to store $(\uu, \vv-\uu)$, which is an integer ($\uu$) and
a bit since $\uu \leq \vv \leq \uu+1$.

Back to our algorithm, each associated-tree $\AT^w$ of the
decomposition of $T_v$ will be described by its vector $vect(w)$, and
the message sent by a node $v$ to its father $v_0$ contains the vector
of all associated-trees of the decomposition. However if the
decomposition does not verify some specific properties, this
information is not sufficient to compute the process number of $T_v$.
It is why we need the notion of hierarchical decomposition.

\subsection{Hierarchical decomposition}

\begin{wrapfigure}[6]{r}{3,8cm}
  \vspace{-0.8cm} \hspace{-1cm}
  \begin{center}
  \hspace{-1cm}\scalebox{0.4}{\input{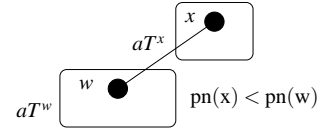}}
   \vspace{-0.3cm}  \hspace{-1cm}
   \caption{$\AT^x < \AT^w$.}
   \label{figx:order}
  \end{center}
\end{wrapfigure}

In a {\it hierarchical decomposition} of $T_v$, we impose that an
associated-tree $\AT^w$ has a process number higher than the
associated-tree $\AT^x$ containing the father of $w$, as illustrated
in Figure~\ref{figx:order}. We also impose that a hierarchical
decomposition has at most one stable associated-tree and if there is
one it has to be minimal according to this order. Finally we impose
that all unstable associated-trees satisfies
Property~\ref{prop:SxStructure}. Figure~\ref{fig:tarbre} gives an
example of a hierarchical decomposition of a tree with process
number~9.



\begin{newproperty}[c.f. {\bf Figure~4}]\label{prop:SxStructure}
  Given a node $w$, its associated-tree $\AT^w$, the subtree $T_w$
  rooted at $w$, and $\Gamma(w) \cap T_w = \{w_1,\dots,w_{k}\}$, if
  $\AT^w$, and so $w$, is unstable it has the following structure: $w$
  has two neighbours $w_1,w_2 \in \Gamma(w) \cap T_w$ which are
  stables and such that $\uu(w_1)=\uu(w_2)=\uu(w)$. Furthermore
  $\AT^w$ is formed by its root $w$, the two stable associated-trees
  $\AT^{w_1}$ and $\AT^{w_2}$ and of $l\leq k-2$ other subtrees
  $T^{w_3},\dots, T^{w_{l+2}}$ whose roots are visited neighbours and
  whose process number is at most $\uu(w)-1$. Notice that the subtrees
  $T^{w_3},\dots, T^{w_{l+2}}$ are not necessarily the
  associated-trees $\AT^{w_3},\dots, \AT^{w_{l+2}}$.
\end{newproperty}

\begin{figure}[t]
  \begin{center}
    \scalebox{0.38}{\input{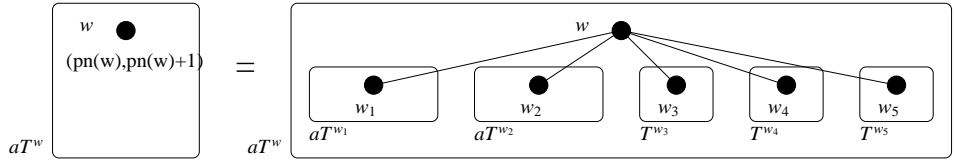}}\vspace{-0.1cm}
    \caption{Structure of an unstable associated-tree $\AT^w$.
      $vect(w_1)=vect(w_2)=(\uu(w),\uu(w))$ and $\forall
      i\in\left[3,5\right], \pn(T^{w_i})<\uu(w)$.}
  \end{center}
  \label{fig:SxStructure}
\end{figure}

\begin{figure}[tb]
  \begin{center}
    \scalebox{0.35}{\input{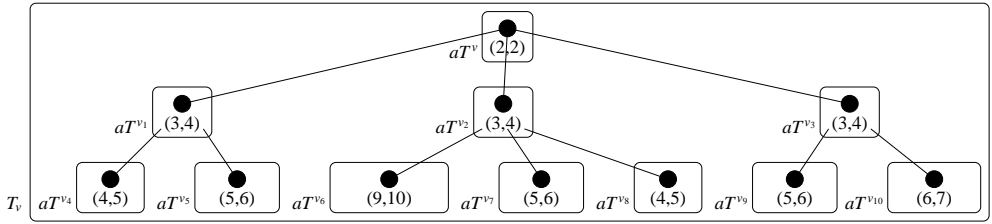}}\vspace{-0.1cm}
    \caption{Example of a hierarchical decomposition of a tree $T_v$
      with process number 9.}
    \label{fig:tarbre}
  \end{center}
  \vspace{-3mm}
\end{figure}

To describe a given hierarchical decomposition, a node $v$ stores a
vector and a table encoding the shape of the associated-trees $\AT^v$.
We will see with Theorem~\ref{thm:carac} that it is sufficient to
compute the process number of $T_v$. More precisely $v$ stores:
\begin{itemize}
\item The vector of the stable associated-tree of the decomposition if
  there is one, $(-1,-1)$ otherwise;
\item A table $t_v$ of length $L(t_v)=\max_{w \in R_v}(\uu(w))$ which
  in cell $i$, noted $t_v[i]$, contains the number of unstable
  associated-trees whose vector is $(i,i+1)$ in the
  decomposition. (Remember that (1,2) is considered as stable, hence
  the first cell always contains 0).
\end{itemize}

For example in Figure~\ref{fig:tarbre}, $v$ and $v_1$ store respectively:
\begin{center}
  \begin{tabular}{l@{: }l}
    $HD(v)$ &  $t_{v}=$\ 
    \begin{tabular}{|c|c|c|c|c|c|c|c|c|}
      \hline
      0 & 0 & 3 & 2 & 3 & 1 & 0 & 0 & 1\\
      \hline
    \end{tabular}
    \ and $(\uu(v),\vv(v))=(2,2)$\\
    $HD(v_1)$ & $t_{v_1}=$\ 
    \begin{tabular}{|c|c|c|c|c|c|c|c|c|}
      \hline
      0 & 0 & 1 & 1 & 1\\
      \hline
    \end{tabular}
    \ and $(\uu(v_1),\vv(v_1))=(-1,-1)$
  \end{tabular}
\end{center}

\begin{lemma}\label{lem:utile}
  Let $T = (V,E)$ be a tree rooted at $r$ and $\AT^w$, $r \notin \AT^w$, an unstable associated-tree
  rooted at $w \in V$ in a hierarchical decomposition. If $\pn(\AT^w)
  = p$, $\pn(T)=p$ iff $\pn(T \setminus \AT^{w}) \leq p-1$.

  Furthermore if $\pn(T)=p$, $T$ is unstable.
\end{lemma}

\begin{proof}
  If there is a tree $\AT^x$ in the hierarchical decomposition with
  $\pn(\AT^x) > p$ then $\pn(T \setminus \AT^{w}) > p$. From now on we
  assume that for all $\AT^x$ of the hierarchical decomposition,
  $\pn(\AT^x) \leq p$. Using the properties of a hierarchical
  decomposition, it implies that $w$ is the only node through which
  $\AT^{w}$ is connected to the rest of $T$.

  By Property~\ref{prop:SxStructure}, $\AT^{w}$ is formed by its root
  $w$, two stable subtrees $T^{w_1}$ and $T^{w_2}$ with process number
  $p$ and some other subtrees with process number less than $p-1$.

  If $T \setminus \AT^w$ has process number at least $p$ then $w$ is a
  node with three branches having process number at least $p$. Hence,
  by Theorem~\ref{p3}, $T$ has process number at least $p+1$.

  Otherwise $\pn(T \setminus \AT^w) < p$ and we describe a $p$-process
  strategy.  We start by an optimal process strategy the stable
  associated-tree $\AT^{w_1}$. It uses $p$ agents and finishes with
  $w_1$ occupied by an agent.  Then we place an agent on $w$ and
  process $w_1$. We continue with an optimal process strategy of $T^w
  \setminus \AT^{w_2}$, it uses at most $p-1$ extra agents.

  Now, since $\pn(T \setminus \AT^w) < p$, we continue with a
  $(p-1)$-process strategy of $T \setminus \AT^w$. We then place an
  agent on $w_2$ and process $w$. It now only remains to process
  $\AT^{w_2}$ starting at $w_2$ which can be done with $p$ agents by
  assumption.

  $T$ is clearly unstable since it contains an unstable subtree $\AT^w$ with same process number which does not contain the root of $T$.
\end{proof}

\begin{theorem}\label{thm:carac}
  Given a rooted tree $T$, a table $t$ and a vector $vect=(\uu,\vv)$,
  if there is a hierarchical decomposition of $T$ described by
  $(vect,t)$, we can compute $\pn(T)$. More precisely: \vspace{-5pt}
  \begin{itemize}
  \item[a)] $\pn(T) = L(t) \Leftrightarrow \exists i \in \left[1..L(t)
    \right]$ such that $t[i] = 0$ and $\forall j \in \left[ i+1..L(t)
    \right]$\ $t[j] = 1$. Furthermore $T$ is unstable.

  \item [b)]If $\pn(T) \neq L(t)$ then $\pn(T) = \max\set{\uu,L(t)+1}$ and
    $T$ is stable.
  \end{itemize}
\end{theorem}

The Property a) means that if in the table $t$ of a hierarchical
decomposition there is a cell with a 0 followed only by cells full of
1, then the process number of a tree accepting such a hierarchical
decomposition has process number $L(t)$.

\vspace{2mm}
\begin{proof}[of Theorem~\ref{thm:carac}]
  First remark that the process number is at most $L(t)+1$.

  \noindent By induction on $L(t)$.\vspace{-5pt}
  \begin{itemize}
  \item If $L(t) = 0$, $T$ is a single node and $\pn(T)=0$. If $L(t) =
    1$, $T$ is a stable tree with vector (1,1) or (1,2). In both case
    $\pn(T)=1$. If $L(t) = 2$ and $t[2]=0$, $T$ is a stable tree with
    vector (2,2) and $\pn(T)=2$. If $t[i]=0$ for all $ i\leq L(t)$,
    $T$ is a stable tree with vector $(L(t), L(t))$ and $\pn(T)=
    L(t)$.
  \item When $L(t) \geq 2$ and $t[L(t)]=1$. We call the
    associated-tree of the hierarchical decomposition having process
    number $L(t)$ $\AT^{w}$ and $w$ its root. By Lemma
    \ref{lem:utile}, $\pn(T)=L(t) \Leftrightarrow \pn(T \setminus
    \AT^{w})\leq L(t)-1$.
    \begin{itemize}
    \item If $\exists i \in \left[1..L(t) \right]$ with $t[i] =
      0$ and $\forall j \in \left[ i+1..L(t) \right]$\ $t[j] = 1$, we
      have $\pn(T \setminus \AT^{w}) \leq L(t)-1$.
      \begin{itemize}
      \item Indeed, either $t[L(t)-1]=1$ and $\pn(T \setminus \AT^{w})
        = L(t)-1$ by induction, so $\pn(T)=L(t)$.
      \item Or $t[L(t)-1]=0$. In this case either, we have a table
        with only 0 and we are at an initialisation case: $\pn(T
        \setminus \AT^{w}) = L(t)-1$ or we can delete this last cell,
        the length of the table is then $L(t) - 2$ and we are sure
        that $\pn(T \setminus \AT^{w}) \leq L(t)-1$ by the very first
        remark of the proof. In both cases we have once again
        $\pn(T)=L(t)$.
      \end{itemize}
    \item If in $t$ there is a cell with a number bigger than one
      followed by cells full of one until the last cell, then, by
      induction, $\pn(T \setminus \AT^{w}) = L(t)$ and hence
      $\pn(T)=L(t)+1$.
    \end{itemize}
  \item When $L(t) \geq 2$ and $t[L(t)] \geq 2$, we call one of the
    associated-tree of process number $L(t)$ $\AT^{w}$ and $w$ its
    root.  $\pn(T \setminus \AT^{w}) \geq L(t)$, hence, from Lemma
    \ref{lem:utile} $\pn(T) > L(t)$ which means $\pn(T)=L(t)+1$ by the
    very first remark.
  \end{itemize}
  $T$ stable or unstable follows from Lemma~\ref{lem:utile} and the process strategy we described.
\end{proof}

\vspace{-3mm}
\subsection{Minimal hierarchical decomposition}

\begin{figure}[tb]
  \begin{center}
    \scalebox{0.28}{\input{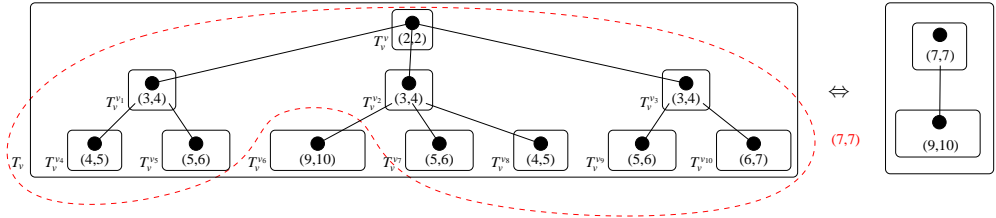}}
    \caption{A simpler hierarchical decomposition of the example of
      Figure \ref{fig:tarbre}.}
    \label{fig:tarbre2}
  \end{center}
  \vspace{-3mm}
\end{figure}

In the example of Figure~\ref{fig:tarbre}, Theorem~\ref{thm:carac}
directly says it has process number 9. If we now consider this example
minus the subtree of vector $(9,10)$, then Theorem~\ref{thm:carac}
says it has process number 7 and furthermore that it is stable.
Hence, we can get another hierarchical decomposition as shown on
Figure~\ref{fig:tarbre2}.

In fact we can generalize this simplification. Given a table $t$ and
an index $i \leq L(t)$, we note $t[1..i]$ the table composed of the
$i$ first cells of $t$. For a given hierarchical decomposition
described by its vector and its table, $HD=(vect, t)$, we call
$HD_i=(vect, t[1..i])$ a \emph{$i$-restricted hierarchical
  decomposition}. Notice that if $HD$ is a hierarchical decomposition
of a tree $T$, then $HD_i$ is a hierarchical decomposition of the
subtree composed of the associated-trees having process number at most
$i$.

A last definition, if a tree accepts several hierarchical
decompositions, we say they are \emph{equivalent}.

We now describe the simplification of a given hierarchical
decomposition $HD=(vect, t)$ of a tree $T$. If there is $i\leq L(t)$
such that a tree $T_i$, whose hierarchical decomposition is described
by $HD_i=(vect, t[1..i])$, has process number $i+1$, then $HD$ is
equivalent to a simpler hierarchical decomposition $HD'=((i+1,i+1),
t')$, where $L(t')=L(t)$, $t'[j]=0$ for $j\leq i+1$, and $t'[j]=t[j]$
for $j>i+1$. If no such $i$ exist, the hierarchical decomposition can
not be simplified.

We call a hierarchical decomposition we can not simplify a
\emph{minimal hierarchical decomposition}. Our algorithm will compute
such decompositions for each subtree $T_v$, $v\in V$. Furthermore we
have:

\begin{lemma}
  \label{lem:binaire}
  Let $HD=((\uu,\vv),t)$ be a minimal hierarchical decomposition. For
  all $i\in [2..L(t)]$, we have $t[i]\in\set{0,1}$.
\end{lemma}


\section{Distributed algorithm for the process number}
\label{sec:algodirect}
We can now describe precisely algorithm \algoHD:

\begin{itemize}
\item The algorithm is initialized at the leaves. Each leaf sends the
  message $((0,0),[\ ])$ (where $[\ ]$ represents a table of length
  $0$) to its only neighbour which becomes its father.

\item A node $v$, which has received messages from all its neighbours
  but one, computes the minimal hierarchical decomposition of $T_v$
  using Algorithm~\ref{alg:fusion}. Then it sends  
  $(\uu(T_v),\vv(T_v)),t_v)$ to its last neighbour, its father.

\item The last node $w$ receives a message from all its neighbours, it
  computes the minimal hierarchical decomposition of $T_v=T$ and
  Theorem~\ref{thm:carac} gives the process number $\pn(T)$.  $w$ is
  called the root of~$T$.
\end{itemize}

\begin{remark}\label{rem:choixracine}
  It may happen that two adjacent nodes $v$ and $w$ receive a message
  from all their neighbors. It is the case when node $v$, after
  sending its message to its last neighbor $w$, receives a message
  from $w$. In this case, both $v$ and $w$ are potential candidates to
  be the root of the tree. There are two possibilities to solve this
  problem. If each node has a unique identifier (e.g. MAC address)
  known by its neighbors, then the one of $v$ and $w$ with the largest
  identifier becomes the root, otherwise, $u$ and $w$ send each other
  a random bit, repeat in case of equality, and the 1 win.
\end{remark}

\begin{algorithm}[htbp]
  \caption{Computation of the minimal hierarchical decomposition}
  \label{alg:fusion}
  \begin{algorithmic}[1]
\REQUIRE{$v_1, ..., v_d$ the visited neighbours of v, and the
  corresponding minimal hierarchical decompositions
  $HD(v_i)=((\uu(v_i),\vv(v_i)),t_{v_i})$}
\REQUIRE{$t_v^{int}$, a table such that $t_v^{int}[i] := t_{v_{1}}[i]
  + ... + t_{v_{d-1}}[i]$, $\forall i\in [2..\max_{1\leq j\leq d}
  L(t_{v_j})]$.}
\REQUIRE{$M_v := \left\{v_{i} ;\ \forall j \in [1..d-1] ,\ \uu(v_{j}) \leq \uu(v_{i}) \right\}$}
\COMMENT{all $v_i$ such that $\uu(v_i)$ is maximum}
\ENSURE{$vect(v)$ and $t_v$}

\COMMENT{computation }

\STATE Let $(p_v,p_v^+)$ be the vector of the associated-tree of $v$
\IF[Initial cases] {$\forall v_{i} \in M_v$, $\pn(v_{i}) < 2$}\label{ln:ln1}
\STATE $(p_v,p_v^+):= \left\{
\begin{array}{ll}
  (0,0) & \mbox{when } \forall v_{i} \in M_v,\ \uu(v_i)=-1\\
  (1,1) & \mbox{when } \forall v_{i} \in M_v,\ \uu(v_i)=0\\
  (1,2) & \mbox{when } |M_v|=1 \mbox{ and } vect(v_i)=(1,1)\\
  (2,2) & \mbox{otherwise}
\end{array}\right.$\label{ln:ln2}
\ELSE[general cases]
	\IF[$v$ is unstable] {$\left|M_v\right| = 2$}
			\STATE $(p_v,p_v^+) := (\pn(v_{i}),\pn(v_{i})+1)$,
                        where $v_{i} \in M_v$
	\ELSE[$v$ is stable]
		\IF[Theorem~\ref{p3}] {$\left|M_v\right| > 2$}
			\STATE $(p_v,p_v^+) :=
                        (\pn(v_{i})+1,\pn(v_{i})+1)$, where $v_{i} \in M_v$
		\ELSE
			\STATE $(p_v,p_v^+) := (\pn(v_{i}),\pn(v_{i}))$,
                        where $v_{i} \in M_v$
		\ENDIF
	\ENDIF
\ENDIF
\vspace{2mm}
\COMMENT{computation of the table}
\STATE $L(t_v) := \max\set{L(t_v^{int}),p_v}$
\STATE $t_v:=t_v^{int}$

\IF{$p_v<p_v^+$ and $p_v > 1$}
	\STATE $t_v[p_v] := t_v[p_v] + 1$
        \STATE $t_v[j]:=0$, $\forall j\in [2..p_v-1]$
	\STATE $(p_v,p_v^+) := (-1,-1)$
\COMMENT{Here, $(p_v,p_v^+)$ is stable}
\ENDIF
\vspace{2mm}

\STATE Let $k$ be such that $t_v[k]>1$ and $t_v[i]\leq 1$, $\forall i\in[k+1..L(t_v)]$
\STATE Let $k_1$ be such that $t_v[k_1]=0$ and $t_v[i]=1$, $\forall i\in[k..k_1-1]$
\IF{$t_v[p_v]=0$}
\STATE $k_2:=p_v$
\ELSE 
\STATE Let $k_2$ be such that  $t_v[k_2]=0$ and $t_v[i]>0$, $\forall i\in[p_v..k_{2}-1]$
\COMMENT{We assume that there exists a virtual cell $t_v[L(t_v)+1]=0$}
\ENDIF
\IF{$k, k_1$ and  $k_2$ exist}
\STATE $t_v[i]:=0$, $\forall i\in[2..\max(k_1,k_2)] := 0$
\STATE $vect(v) := (\max(k_1,k_2),\max(k_1,k_2))$
\ELSE[the hierarchical decomposition is already minimal]
\STATE $vect(v) := (p_v,p_v^+)$
\ENDIF

\end{algorithmic}
\end{algorithm}


\vspace{-3mm}
\begin{lemma}
  Given a tree $T=(V,E)$, with $|V|=n$, the time complexity of
  Algorithm~\ref{alg:fusion} is $O(\log{n})$.
\end{lemma}
\begin{proof}
  All operations are linear in $L(t_v)$, and $L(t_v)\leq \uu(T)\leq
  \log_3{n}$.  
\end{proof}

\vspace{-3mm}
\begin{lemma}
  Given a tree $T=(V,E)$, with $|V|=n$, algo HD computes $\pn(T)$ in
  $n$ steps and overall $O(n\log{n})$ operations.
\end{lemma}
\begin{proof}
  Each node $v$ of degree $d_v$ has to compute $M_v$ (the set of
  neighbors $v_i$ with maximum $\uu(v_i)$) which requires $O(d_v)$
  operations, and $t_v^{sum}$ (the sum of all received tables) that is
  $O(\sum_1^dL(t_v^{sum}))$ operations.  Finally it applies
  Algorithm~\ref{alg:fusion}.  As $\sum_{v\in V}d_v=2(n-1)$, we have
  $\sum_{v\in V}(d_v+\log{n}+\sum_1^dL(t_v^{sum})) =O(n\log{n})$.
\end{proof}

\vspace{-3mm}
\begin{lemma}
  Given a tree $T=(V,E)$, with $|V|=n$, \algoHD sends $n-1$ messages
  each of size $\log_3{n}+2$.
\end{lemma}


\begin{proof}
  Node $v$ sends its minimal hierarchical decomposition to its father,
  that is $HD_v=(vect(v),t_v)$, with $vect(v)=(\uu(v),\vv(v))$.  From
  Theorem~\ref{p3} we know that $L(t_v)\leq \log_3{n}$, from
  Lemma~\ref{lem:binaire}, $t_v$ contains only 0 and 1's, hence we
  need only $\log_3{n}$ bits to transmit $t_v$.  Furthermore, if
  $\uu(v) \geq 1$, $t_v[\uu(v)] = 0$ and $\forall i \leq \uu(v),
  t_v[i] = 0$. Hence we can add an artificial 1 to the cell of $t_v$
  with index $\uu(v)$ to indicate the value $\uu(v)$.

  To summarize, we transmit a table $t$ and two bits $ab$. $ab=00$
  means $vect(v) =(-1,-1)$, $ab=01$ means $vect(v) =(0,0)$, $10$ means
  $vect(v) =(\uu,\uu)$ and $11$ means $vect(v) =(\uu,\uu+1)$. When
  $a=1$, $\uu$ is the index of the first $1$ in the transmitted table
  and $t_v$ is the transmitted table minus this 1. When $a=0$, $t_v$ is
  the transmitted table $t$.
  It is clear that in this coding, each message has size
  $\log_3{n}+2$.
\end{proof}





\vspace{-5mm}

\section{Dynamic and incremental algorithms}
\label{sec:incremental}

In this section, we propose a dynamic algorithm that allows to compute
the process number of the tree resulting of the addition of an edge
between two trees. It also allows to delete any edge. To do this
efficiently, it uses one of the main advantage of the hierarchical
decomposition: the possibility to change the root of the tree without
additional information (Lemma~\ref{lem:changeroot}). From that we
design an incremental algorithm that computes the process number of a
tree.

If we want to join two trees with an edge between their roots then it
is easy to see that Algorithm~\ref{alg:fusion} will do it. However if
we do not join them through the root, a preprocessing to change the
root of the trees needs to be done. In next Section we propose one. To
apply this algorithm, each node needs to store the information
received from each of its neighbors and a table which is the sum of
the received tables: $\forall v_i \in \Gamma(v) \cap T_v: vect_{v_i}$,
$t_{v_i}$ and $t_v^{sum}$. Recall that $t_v^{sum}$ is defined as
$t_v^{sum}[j]=\sum_{v_i \in \Gamma(v) \cap T_v} t_{v_i}[j]$ in the algorithm.

For a given tree $T$, we note $D(T)$ or $D$ if there is no ambiguity
the \emph{diameter} of $T$.

We describe now three functions we will use in the dynamic version of
our algorithm.

\subsection{Functions for updating the process number}

\begin{lemma}[Change of the root] 
  \label{lem:changeroot} 
  Given a tree $T = (V,E)$ rooted at $r_1 \in V$ of diameter $D$, and
  its hierarchical decomposition, we can choose a new root $r_2 \in V$
  and update accordingly the hierarchical decomposition in $O(D)$
  steps of time complexity $O(\log{n})$ each, using $O(D)$ messages of
  size $\log{n}+3$.
\end{lemma}

\begin{proof}  
  We describe an algorithm to change the root from $r_1$ to $r_2$:

  First, $r_2$ sends a message to $r_1$ through the unique path
  between $r_1$ and $r_2$, $r_2=u_0,u_1,u_2,\dots,u_k=r_1$, to notify
  the change.  Then, $r_1$ computes its hierarchical decomposition,
  considering that $u_{k-1}$ is its father. We assume that each
  node $v$ stores the information received from its neighbours and
  $t_v^{sum}$. $r_1$ applies Algorithm~\ref{alg:fusion} using
  all vectors stored but $vect_{u_{k-1}}$ and $t_v^{sum} -
  t_{v_{k-1}}$. Then it sends a message to $u_{k-1}$.

  After, $u_{k-1}$ computes its hierarchical decomposition,
  considering that $u_{k-2}$ is its father, and sends a message to
  $u_{k-2}$. We repeat until $r_2$ receives a message from
  $u_1$. Finally, $r_2$ computes the process number of $T$ and becomes
  the new root. We have a new hierarchical decomposition.

  In this algorithm, $u_i$ substracts the table $t_{u_{i-1}}$
  from $t_{u_i}^{sum}$, and later adds $t_{u_{i+1}}$, computes
  $M_{u_i}$ and finally applies Algorithm~\ref{alg:fusion}.  Clearly,
  all computation requires $O(\log{n})$ operations. The messages need
  one more bit than in the previous algorithm to indicate whether a
  table has to be added or substracted.
\end{proof}


\begin{lemma}[Addition of an edge]\label{lem:addedge}
  Given two trees $T_{r_1} = (V_{1},E_{1})$ and $T_{r_2} =
  (V_{2},E_{2})$ respectively rooted at $r_1$ and $r_2$, we can add
  the edge $(w_1,w_2), w_1 \in V_{1}$ and $w_2 \in V_{2}$ and compute
  the process number of $T = (V_{1} \cup V_{2},E_{1} \cup E_{2} \cup
  (w_1,w_2))$, in at most $D$ steps.
\end{lemma}

\begin{proof}
  First we change the roots of $T_{r_1}$ and $T_{r_2}$ respectively to
  $w_1$ and $w_2$ using Lemma~\ref{lem:changeroot}. Then, $w_1$ and
  $w_2$ decide of a root (see Remark~\ref{rem:choixracine}) which
  finally computes the process number of $T$.
\end{proof}

\begin{lemma}[Deletion of an edge]\label{lem:deleteedge}
  Given a tree $T = (V,E)$ rooted at $r$ and an edge $(w_1,w_2) \in
  E$, after the deletion of edge $(w_1,w_2)$, we can compute the
  process number of the two disconnected trees in at most $D$ steps.
\end{lemma}

\begin{proof}
  W.l.o.g. we may assume that $w_2$ is the father of $w_1$. Let
  $T_{w_1}$ be the subtree rooted at $w_1$ and $T \setminus T_{w_1}$
  the tree rooted at $r$. Remark that it includes $w_2$.  The process
  number of $T_{w_1}$ is deduced from the previously computed
  hierarchical decomposition.
  Now, to compute the process number of $T \setminus T_{w_1}$, we
  apply the change root algorithm and node $w_2$ becomes the new root
  of $T \setminus T_{w_1}$.
\end{proof}

\subsection{Incremental algorithm}

From Lemma~\ref{lem:addedge}, we obtain an incremental algorithm
(\incHD) that, starting from a forest of $n$ disconnected vertices
with hierarchical decomposition $((0,0,)[\ ])$, add tree edges one by
one in any order and updates the process number of each connected
component. At the end, we obtain the process number of $T$.

This algorithm is difficult to analyze in average, but the best and
worst cases are straightforward:
\begin{itemize}
\item Worst case: $T$ consists of two subtrees of size $n/3$ and
  process number $\log_3(n/3)$ linked via a path of length
  $n/3$. Edges are inserted alternatively in each opposite
  subtrees. Thus \incHD requires $O(n^2)$ steps and messages, and
  overall $O(n^2\log{n})$ operations
\item Best case: edges are inserted in the order induced by \algoHD
  (inverse order of a breadth first search). \incHD needs $O(n)$
  messages and an overall of $O(n\log{n})$ operations.
\end{itemize}

Actually, the overall number of messages is $O(nD)$ and the number of
operations is $O(nD\uu(T))$. They both strongly dependent on the order
of insertion of the edges.  Thus an interesting question is to
determined the average number of messages and operations.




\section{Improvements and extensions}
\label{sec:discussion}

\paragraph{Reducing the amount of transmitted information}
In our algorithms, it is possible to reduce the size of some messages
and so the overall amount of information transmitted during the
algorithm.  For example, instead of transmitting $\log{n}$ bits for
$t$, we may transmit only $L(t)$ bits plus the value $L(t)$ on
$\log\log{n}$ bits.  Overall we will exchange less than
$n(\uu(T)+\log_2\log_3{n}+2+\varepsilon)$ bits, where $\varepsilon=1$
for the dynamic version of the algorithm (\incHD).  Further
improvements are possible with respect to the following lemma.

%
%

\begin{lemma}
  Assuming that when an edge is added at vertex $v$, $v$ asks its
  neighbours information once and simultaneously, any dynamic
  algorithm satisfying this assumption induces a transmission of at
  least $\frac{k-1}{k}n(\pn(T)-2)$ bits for any $k\in \N$ and value of
  $\pn(T) \leq \log_3(n/k)$ in some trees $T$.

\end{lemma}

\begin{proof}
  Suppose that we are given a dynamic algorithm such that when an edge
  is added at vertex $v$, $v$ asks its neighbours information once and
  simultaneously, and let $k>1$ be an integer.  We consider a tree
  made of a path $u$-$v$ of length $\frac{k-1}{k}n$ with a tree $T'$
  at $u$.  One of the messages received by $v$ gives information about
  $T'$.  If for all tree $T'$ with process number $p$, the algorithm
  uses less than $p-2$ bits to encode this message, and since there is
  more than $2^{p-2}$ hierarchical decompositions corresponding to a
  tree with process number $p$, there exists two trees $T'_1$ and
  $T'_2$ with different minimal hierarchical decompositions but which
  are encoded in the same way.  We note $T_1$ when $T'=T'_1$ and $T_2$
  when $T'=T'_2$. Then, it exists a tree $T"$ such that if we join it
  to (w.l.o.g) $T_1$ at $v$, the process number of $T_1$ increases by
  one whereas if we join $T"$ to $T_2$ at $v$, the process number of
  $T_2$ does not increase.

  Hence, there is a tree $T'$ for which the algorithm encodes the
  information transmitted to $v$ on at least $p-2$ bits. For this $T'$
  in our construction of $T$, the information received by $v$ comes
  from $u$ and hence it has transited through $\frac{k-1}{k}n$ nodes.
  Therefore, the total of transmitted bits is at least
  $\frac{k-1}{k}n(p-2)$.
\end{proof}

\begin{corollary}
  Assuming that when an edge is added at vertex $v$, $v$ asks its
  neighbours information once and simultaneously, any dynamic
  algorithm induces a transmition of at least
  $\frac{k-1}{k}n(\log_3{n})$ bits in some large enough trees, for any
  $k\in \N$.

\end{corollary}

\begin{proof}
  Let $k \in \N$. By the previous Lemma for $k+1$, there is a tree $T$
  with process number $\log_3(n/(k+1))$ which induces a transmition of
  at least $\frac{k}{k+1}n(\log_3(n/(k+1))-2)$ bits, and this larger
  than $\frac{k-1}{k}n(\log_3{n})$ when $\log{n}>k^2(\log_3(k+1)+2)$.
\end{proof}

%

\paragraph{Reducing the number of operations}
It makes no doubt that the worst case complexity of \incHD and more
specifically of Lemma~\ref{lem:addedge} can be seriously improved. In
particular, instead of changing the roots of both trees, we may change
only $r_1$ to $w_1$, then transmit information in the direction of
$r_2$, and eventually stop the transmissions before $r_2$ if the
minimal hierarchical decomposition of some node remains unchanged.

It is also interesting to notice that using arguments similar
to~\cite{EST94}, we can get a centralized algorithm using a linear
number of operations.

\paragraph{Trees and forests of unknown size}
If the size $n$ of the tree is unknown, a node encodes each bit of the
transmitted table $t$ on 2 bits, that is 00 for 0 and 01 for 1. It
allows to use 11 to code the end of the table and hence to know its
length. Thus the receiver may decode the
information without knowing $n$. In this coding the table requires
$2L(t)+2$ bits and the transmission requires $2L(t)+4+\varepsilon$
bits, where $\varepsilon=1$ for $\incHD$ and 0 for $\algoHD$. Remember
that $L(t)\leq \uu(T)$.

\paragraph{Computing other parameters}
Our algorithms can be adapted to compute the node search number or the
pathwidth of any tree with the same time complexity and transmission
of information. For that, it is sufficient to change the values of the
initial cases (lines~\ref{ln:ln1} and~\ref{ln:ln2}) in
Algorithm~\ref{alg:fusion}.

For the node search number we would use the initial cases of the left
of Figure~\ref{fig:newsmallcases}. Notice that in this case we do not
use the vector $(1,2)$.

\begin{figure}[htb]
  \hspace{-1em}\begin{minipage}[t]{.55\textwidth}
    \begin{algorithmic}[l]
 \small     \IF{$\forall v_{i} \in M_v$, $\pn(v_{i}) < 2$}
      \STATE $(p_v,p_v^+):= \left\{
      \begin{array}{ll}
        (1,1) & \mbox{when } \forall v_{i} \in M_v,\ \uu(v_i)=-1\\
        (2,2) & \mbox{otherwise}
      \end{array}\right.$
    \ENDIF
  \end{algorithmic}
\end{minipage}
\hfill
\begin{minipage}[t]{.45\textwidth}
  \begin{algorithmic}[l]
\small    \IF{$\forall v_{i} \in \Gamma(v)$, $\vv(v_{i}) < 2$}
    \STATE $(p_v,p_v^+):= \left\{
    \begin{array}{ll}
      (0,0) & \mbox{when } |M_v|=0\\
      (1,1) & \mbox{when } |M_v|=1\\
      (1,2) & \mbox{when } |M_v|=2\\
      (2,2) & \mbox{otherwise}
    \end{array}\right.$
  \ENDIF
\end{algorithmic}
\end{minipage}
\label{fig:newsmallcases}
\caption{Initial cases for node search number (left) and edge search
  number (right).}
\end{figure}

For the edge search number of a tree, we can prove that $\ns(T)-1 \leq
\es(T) \leq \ns(T)$, whereas on a general graph we only have $\ns(T)-1
\leq \es(T) \leq \ns(T)+1$. To adapt Algorithm~\ref{alg:fusion} for
the edge search number, we would use the initial cases of the right of
Figure~\ref{fig:newsmallcases} plus the extra rule that all received
vectors $(1,2)$ are interpreted as if they were vectors $(2,2)$. Also,
if all received vectors verifies $\vv(v_{i}) < 2$, $M_v$ is the set of
all received vectors different from (-1,-1). Notice that it gives the
first algorithm to compute the edge search number of trees.

Algorithm \algoHD has been implemented for the process number, the
node search number and the edge search number, as well as
corresponding search strategies~\cite{capture}.

\paragraph{About the difference of the parameters}
Finally, the following lemma characterizes the trees for which the
process number (resp. edge search number) equals the pathwidth.
\begin{lemma}
  Given a tree $T$, $\uuu(T)=\pw(T)+1=p+1$ (resp.
  $\uuu(T)=\es(T)+1=p+1$) iff there is a node $v$ such that any
  components of $T-\{v\}$ has pathwidth at most $p$ and there is at
  least three components with process number (resp. edge search
  number) $p$ of which at most two have pathwidth $p$.
\end{lemma}

This lemma means that the difference between, e.g., the process number
and the pathwidth comes from the difference on trees with smaller
parameter and ultimately from trees with those parameters equal to~1
or~2.


To give such characterisations for more general classes of graphs
remains a challenging problem.

\subsection*{Acknowledgments}
We would like to thanks Nicolas Nisse and Herv\'e Rivano for fruitfull
discussions on this problem.

\bibliographystyle{plain}
\bibliography{bibliopathwidth,AlgotelPathwidth}

\begin{thebibliography}{10}

\bibitem{capture}
\newblock \url{http://www-sop.inria.fr/members/Dorian.Mazauric/
  Capture/index.php.htm}.

\bibitem{CPPS05}
D.~Coudert, S.~Perennes, Q.-C. Pham, and J.-S. Sereni.
\newblock Rerouting requests in wdm networks.
\newblock In {\em AlgoTel'05}, pages 17--20, Presqu'\^ile de Giens, France, mai
  2005.

\bibitem{CoSe07}
D.~Coudert and J-S. Sereni.
\newblock Characterization of graphs and digraphs with small process number.
\newblock Research Report 6285, INRIA, September 2007.

\bibitem{DPS02}
J.~D\'iaz, J.~Petit, and M.~Serna.
\newblock A survey on graph layout problems.
\newblock {\em ACM Computing Surveys}, 34(3):313--356, 2002.

\bibitem{EST94}
J.A. Ellis, I.H. Sudborough, and J.S. Turner.
\newblock The vertex separation and search number of a graph.
\newblock {\em Information and Computation}, 113(1):50--79, 1994.

\bibitem{FT}
F.~V. Fomin and D.~Thilikos.
\newblock An annotated bibliography on guaranteed graph searching.
\newblock {\em Theoretical Computer Science, Special Issue on Graph Searching},
  2008, to appear.

\bibitem{Kin92}
N.~G. Kinnersley.
\newblock The vertex separation number of a graph equals its pathwidth.
\newblock {\em Inform. Process. Lett.}, 42(6):345--350, 1992.

\bibitem{KP86}
M.~Kirousis and C.H. Papadimitriou.
\newblock Searching and pebbling.
\newblock {\em Theor. Comput. Sci.}, 47(2):205--218, 1986.

\bibitem{MHG+88}
N.~Megiddo, S.~L. Hakimi, M.~R. Garey, D.~S. Johnson, and C.~H. Papadimitriou.
\newblock The complexity of searching a graph.
\newblock {\em J. Assoc. Comput. Mach.}, 35(1):18--44, 1988.

\bibitem{Par76}
T.~D. Parsons.
\newblock Pursuit-evasion in a graph.
\newblock In {\em Theory and applications of graphs}, pages 426--441. Lecture
  Notes in Math., Vol. 642. Springer, Berlin, 1978.

\bibitem{RoSe83}
N.~Robertson and P.~D. Seymour.
\newblock Graph minors. {I}. {E}xcluding a forest.
\newblock {\em J. Combin. Theory Ser. B}, 35(1):39--61, 1983.

\bibitem{Sch90}
P.~Scheffler.
\newblock A linear algorithm for the pathwidth of trees.
\newblock In R.~Henn R.~Bodendiek, editor, {\em Topics in Combinatorics and
  Graph Theory}, pages 613--620. Physica-Verlag Heidelberg, 1990.

\bibitem{Sko03}
K.~Skodinis.
\newblock Construction of linear tree-layouts which are optimal with respect to
  vertex separation in linear time.
\newblock {\em J. Algorithms}, 47(1):40--59, 2003.

\end{thebibliography}

\end{document}